\documentclass[lettersize,journal]{IEEEtran}
\usepackage{amsmath,amsfonts}
\usepackage{algorithmic}
\usepackage{algorithm}
\usepackage{array}
\usepackage[caption=false,font=normalsize,labelfont=sf,textfont=sf]{subfig}
\usepackage{textcomp}
\usepackage{stfloats}
\usepackage{url}
\usepackage{verbatim}
\usepackage{graphicx}
\usepackage{cite}
\usepackage{xcolor}
\usepackage{soul}
\usepackage{multirow}
\begin{document}

\title{Multi-Attribute Attention Network for Interpretable Diagnosis of Thyroid Nodules in Ultrasound Images}

\author{
\parbox{\linewidth}{\centering
Van T. Manh, Jianqiao Zhou, Xiaohong Jia, Zehui Lin, Wenwen Xu, Zihan Mei, Yijie Dong, Xin Yang,\\ Ruobing Huang, Dong Ni%
}

% Van T. Manh, Jianqiao Zhou, Xiaohong Jia, Zehui Lin, Wenwen Xu, Zihan Mei, Yijie Dong, Xin Yang, \\ Ruobing Huang, Dong Ni,~\IEEEmembership{Member,~IEEE}

\thanks{Van T. Manh and Jianqiao Zhou contributed equally to this work.}
\thanks{Corresponding author: Ruobing Huang (email: ruobing.huang@szu.edu.cn) and Dong Ni (nidong@szu.edu.cn). }
% \thanks{This work was supported by National Key R\&D Program of China No.2019YFC$\backslash$ 0118300); Shenzhen Peacock Plan (No. KQTD 2016053112051497, KQJSCX2018$\backslash$ 0328095606003); Medical Scientific Research Foundation of Guangdong Province, China (No. B2018031).}
\thanks{This study was supported by the National Natural Science Foundation of China (No. 62171290) and Shenzhen-Hong Kong Joint Research Program (No. SGDX20201103095613036).} 
\thanks{Van T. Manh, Zehui Lin, Xin Yang, Ruobing Huang, Dong Ni are with Medical UltraSound Image Computing (MUSIC) Lab, School of Biomedical Engineering, Health Center, Shenzhen University, China.}
\thanks{Jianqiao Zhou, Xiaohong Jia, Wenwen Xu, Zihan Mei, and Yijie Dong are with the Department of Ultrasound Medicine, Ruijin Hospital, School of Medicine, Shanghai Jiaotong University, China.}
}
% The paper headers
\markboth{Journal of \LaTeX\ Class Files,~Vol.~14, No.~8, ~2022}%
{Shell \MakeLowercase{\textit{et al.}}: A Sample Article Using IEEEtran.cls for IEEE Journals}

% \IEEEpubid{0000--0000/00\$00.00~\copyright~2021 IEEE}
% Remember, if you use this you must call \IEEEpubidadjcol in the second
% column for its text to clear the IEEEpubid mark.

\maketitle

\begin{abstract}
% Accurate identification of nodule malignancy in ultrasound is an important yet challenging task that can elude less-experienced clinicians.  
Ultrasound (US) is the primary imaging technique for the diagnosis of thyroid cancer. However, accurate identification of nodule malignancy is a challenging task that can elude less-experienced clinicians. Recently, many computer-aided diagnosis (CAD) systems have been proposed to assist this process. However, most of them do not provide the reasoning of their classification process, which may jeopardize their credibility in practical use. To overcome this, we propose a novel deep learning framework called multi-attribute attention network (MAA-Net) that is designed to mimic the clinical diagnosis process. The proposed model learns to predict nodular attributes and infer their malignancy based on these clinically-relevant features. A multi-attention scheme is adopted to generate customized attention to improve each task and malignancy diagnosis. Furthermore, MAA-Net utilizes nodule delineations as nodules spatial prior guidance for the training rather than cropping the nodules with additional models or human interventions to prevent losing the context information. Validation experiments were performed on a large and challenging dataset containing 4554 patients. Results show that the proposed method outperformed other state-of-the-art methods and provides interpretable predictions that may better suit clinical needs.
\end{abstract}


\section{Introduction}
\label{sec:introduction}
\IEEEPARstart{T}{hyroid} nodules are a common pathology with an incidence rate of 19 to 68\% \cite{haugen20162015}, and there has been a 3- to 15-fold increase between 1993 and 2011 \cite{schnadig2018overdiagnosis}. Today, thyroid cancer is listed as the seventh most common cancer in women and the fifteenth most common cancer in men \cite{papini2002risk,rahbari2010thyroid}.%It is especially common for women around 20-55 years old. 

\begin{figure*}
    \centering
    \includegraphics[width = \linewidth]{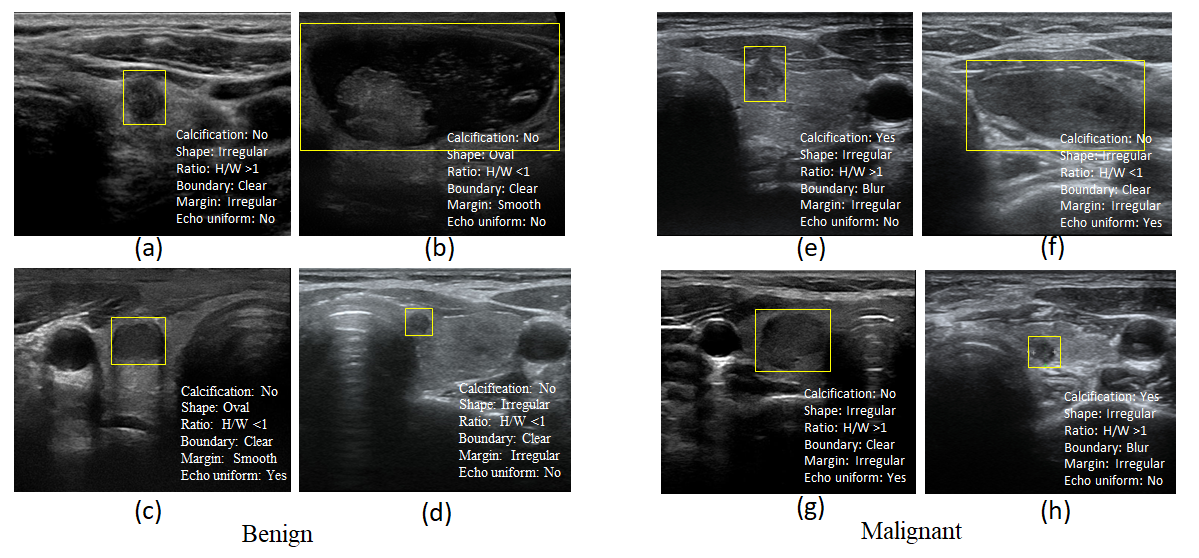}
    \caption{Typical examples of benign and malignant thyroid nodules. The yellow rectangles indicate the sizes and the positions of the nodules. The corresponding nodular attributes are also listed in white. }
    \label{fig1}
\end{figure*}

Ultrasonography (US) is the preferred tool for the diagnosis of thyroid nodules as it is non-invasive, real-time, and non-ionizing. However, accurate identification of the nodule malignancy is challenging and requires expertise. Fig. \ref{fig1} shows some typical examples of benign and malignant nodules.  The sizes and the appearances of the benign cases can be very different (e.g. Fig. \ref{fig1} (b), (d)), indicating large intra-class variance. Meanwhile, some malignant nodules are visually similar to benign cases, suggesting a small inter-class difference.

To overcome this challenge, the Thyroid Imaging Reporting and Data System (TI-RADS) \cite{grant2015thyroid} was created to advise clinical practice. It recommends the following diagnosis process:  clinicians should first characterize the nodules using attributes such as the shape, ratio, boundary, margin, echo uniformity, and calcification. These nodular characteristics have been associated with a higher likelihood of malignancy.  Based on the outcome of these attributes, clinicians estimate the probability of malignancy accordingly and decide whether a fine-needle aspiration (FNA) biopsy is necessary.        

On the other hand, the practical implementation of this guideline may suffer from subjectivity while its accuracy also depends on the experience and the skill of clinicians. For example, it can be difficult to define the two cases in Fig. \ref{fig1} (a), (e) whether their boundaries are blurry or clear, and if their shapes are circular or irregular. As a result, high inter-observer variability has been reported in various studies in classifying these attributes \cite{choi2010interobserver},\cite{hoang2018interobserver}. However, inaccurate recognition of these US characteristics can lead to over-diagnosis (e.g. unnecessary FNA biopsy) that may bring harm to the patients \cite{schnadig2018overdiagnosis}.

To solve this, many computer-aided diagnosis (CAD) systems have been developed. Overall, these methods have achieved promising performances while they usually involve piece-wise models that could be accompanied by error propagation. In specific, they first detect or segment the nodule to crop the image, then use another model to perform classification (e.g  \cite{liu2019automated, buda2019management}). Note that the errors that occur in the localization or the segmentation stage can lead to inaccurate cropping of the images that ultimately hampers the final classification accuracy. Some methods therefore \cite{guan2019deep},\cite{chi2017thyroid} require manual selection of nodule regions before analysis, while it may add additional clinical burdens. Furthermore, most state-of-the-arts CAD systems are CNN-based \cite{liu2019automated},\cite{guan2019deep},\cite{chi2017thyroid}, which may be difficult to interpret and comprehend.

In this work, we propose a novel deep learning-based framework called multi-attribute attention network (MAA-Net). The main contributions are:
 
\begin{itemize}
    \item An end-to-end approach. The framework takes the original US images as input and outputs malignancy of the nodules as well as various nodular attributes. It does not require additional models to pre-localize the nodules or pre-processing to calibrate the image appearance.      
    \item An interpretable CAD system. The model mimics the clinical diagnosis process by making predictions based on predicted nodular attributes. Given this information, the clinicians can better understand the reasoning of the AI CAD system and utilize its predictions to assist diagnosis.
    \item Proposing an attribute-based multi-attention scheme. It can guide the model's attention to the different regions of interest to learn specific features concerning each attribute. 
    \item Introducing a location-aware attention map to incorporate the prior spatial information while avoiding cutting or cropping images stage, which is burdensome and could deplete the potential  context information. As a by-product, this method enables automated generation of location heatmaps of the nodules during testing. Experiment results proved that these heatmaps correlate well with the ground truth delineations. 

\end{itemize}

The rest of the paper is arranged as follows. In Section II and III, we discuss the related works and introduce the proposed method. Section IV and V report the validation experiments and results. Finally, we conclude the paper in Section VI.

\section{Related Works}
\subsection{CAD Systems for Thyroid Cancer Diagnosis}
There has been a lot of research on developing CAD systems for thyroid cancer diagnosis. Most of the traditional approaches combine feature extraction methods with classical machine learning classifiers. For example, Keramidas \emph{et al.}~\cite{keramidas2008thyroid} extracted fuzzy local binary patterns as noise-resistant textural features and used the support vector machine (SVM) as the classifier. Similarily, Ardakani \emph{et al.}~\cite{ardakani2018predictive} also used SVM to classify 49 morphological or textural features extracted from nodule images. Tsantis \emph{et al.}  \cite{tsantis2009morphological} calculated the wavelet local maxima and the shape features to characterize the segmented nodules. Singh \emph{et al.}  \cite{singh2012ultra} used the gray level co-occurrence matrix features to construct a k-nearest neighbor (KNN) model for thyroid nodule classification. Acharya \emph{et al.}~\cite{acharya2014computer} utilized gray-scale features calculated based on stationary wavelet transform. They also compared the performance of several commonly-used classifiers. Raghavendra \emph{et al.}  \cite{raghavendra2017fusion} used the fusion of Spatial Gray Level Dependence Features (SGLDF) and fractal textures to decipher the intrinsic nodular structure. Although of the efforts, as thyroid nodules vary in location, shape, size, and internal characteristics, these hand-crafted features may have difficulty in accommodating all scenarios and differentiating the malignant nodule from the benign ones.

Recently, deep learning (DL) based methods have been adopted to design similar CAD systems. Some of them require human-intervention or pre-processing before analyzing. For example, in Ma \emph{et al.}~\cite{ma2017cascade}, they required radiologists to select the regions of interest (ROIs) in original US images. They then used a cascaded CNN model to classify these ROIs.  Similarly, Guan \emph{et al.} cropped the nodule region based on manual annotation and used the cropped images to train an Inception-V3 model \cite{guan2019deep}. In Chi \emph{et al.}~\cite{chi2017thyroid} , all US images are pre-processed to calibrate their scanning scale and remove possible artifacts. The authors then fine-tuned a GoogLeNet model as a feature extractor to train a Random Forest classifier. These methods could add clinical burden or may have difficulty in adapting to new datasets with different image appearances. 

Others, therefore, turn to fully automated approaches by adding an additional localization stage to remove the background regions. For example, Liu \emph{et al.} \cite{liu2019automated} proposed a two-stage CAD system, which used a multi-scale region-based detection network for nodule detection and a multi-branch classification network to capture and enhance group characteristics for the final malignancy classification . Buda \emph{et al.} \cite{buda2019management} first detected the nodules using the Faster-RCNN model and cropped the US images to the nodule area. The cropped images are then fed into the second CNN network to classify whether the nodules are benign or malignant. Despite exhibiting good performance, these two-staged methods take risk of depleting potential context information and error propagation from the detection stage to the classification stage.  In other words, incorrect detection results will propagate the errors to the classification stage and result in incorrect or unreliable predictions. Therefore, an end-to-end approach is desired, to can handle the origninal US images directly.

\subsection{Interpretable Diagnosis Methods}

Note that most of the aforementioned methods only offer the malignancy prediction result and fail to explain how the decision was made \cite{elshawi2019interpretability}. It can cause confusion and misunderstanding when applied in a real clinical scenario, which may block a wider adoption of DL algorithms. There has been an increasing need for interpretable deep learning diagnosis systems to overcome this dilemma \cite{carvalho2019machine}. 

Beside the main task, many methods try to combine different information from different tasks to improve the performance and generalizable ability of the model \cite{zhang2021survey}.  The joint training can help the model learn the specific features  of the tasks and share the common features representation to reduce the overfitting. For example, Wang \emph{et al.} \cite{wang2018simultaneous} modified the structure of U-Net by adding a classification branch for classification and segmentation of bone surfaces. Murugesan \emph{et al.} \cite{murugesan2019psi} proposed a deep model to predict  the shape and boundary information for optic cup and disc segmentation. Sbhankar Roy \emph{et al.} \cite{roy2020deep} developed a model to simultaneously predict the COVID-19 disease severity score and provides location of pathological artefacts . These methods are not only help improving the tasks together but also provide the auxiliary predictions to assist the diagnosis.

Considering the TI-RADS attributes are widely used to translate the diagnosis to human-understandable characteristics. We incorporate the learning and prediction of TI-RADS attributes into our model design, which adds the model interpretability naturally.

\begin{figure*}
    \centering
    \includegraphics[width = 0.98\textwidth]{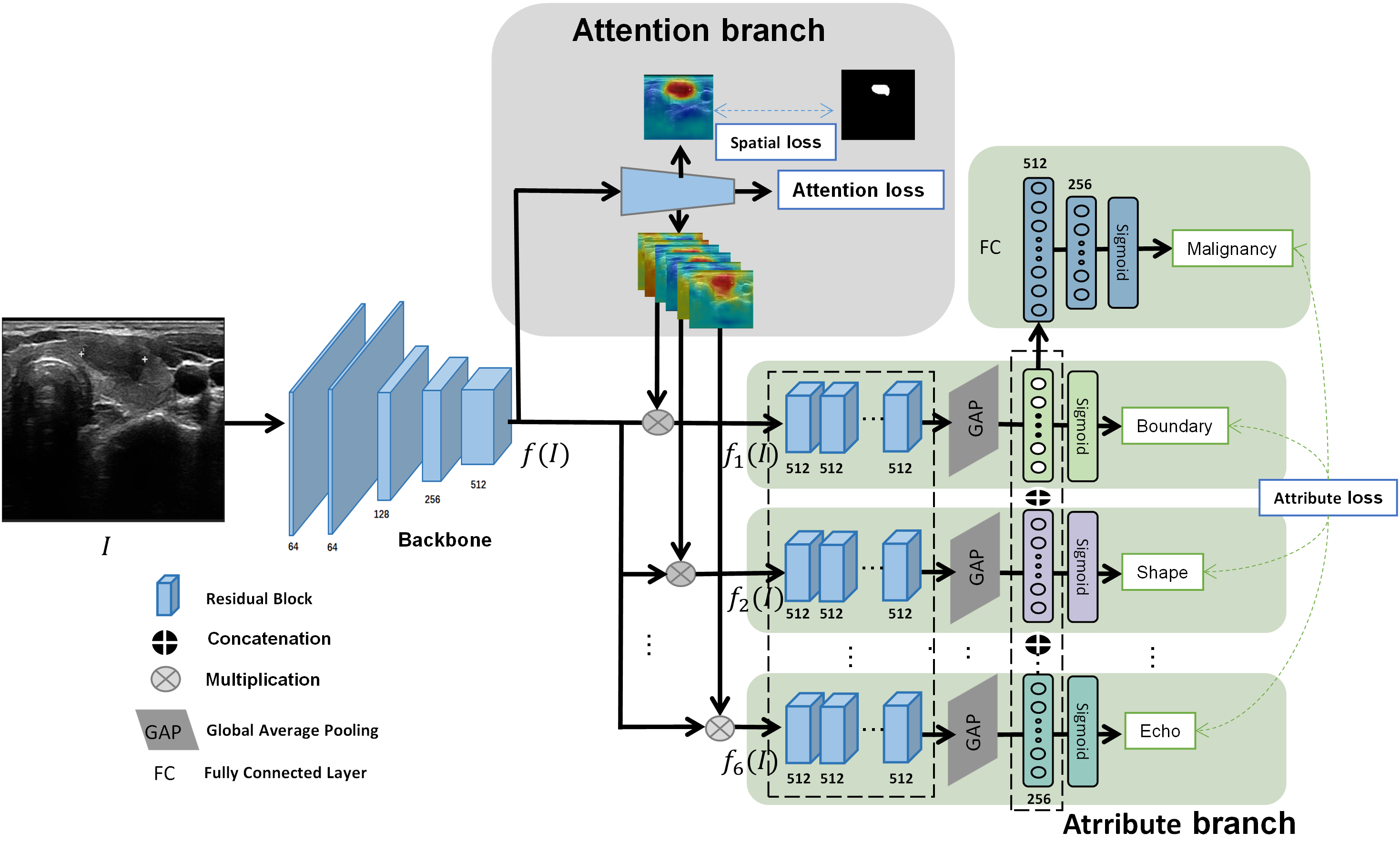}
    \caption{Schematic of the proposed framework. Residual blocks are shown as blue cubes. Different attribute streams are represented using different colors (e.g. green, purple, teal, etc.).}
    \label{fig2}
\end{figure*}

\subsection{Attention-based CNN Models}

The visual attention-based method allows a deep model to focus on important regions of an image adaptively. To provide visual attention maps for deep neural networks, many interesting approaches have been proposed in \cite{zhou2016learning},\cite{selvaraju2017grad}, which are mainly two types: gradient-based visual explanation and activation-based visual explanation. The former uses gradients of a target concept to produce a heatmap of regional importance, while the latter uses the response of a feed-forward propagation. Local interpretable model-agnostic explanations (LIME) is a technique to explore the behavior of the models from variously deformed input to provide visualization \cite{ribeiro2016should}. Visual attention-based CNNs have also been applied in medical image processing. For example, Ouyang \emph{et al.} proposed a dual-sampling attention network to classify the COVID-19 and CAP infection \cite{ouyang2020dual}. They proposed an online module to utilize the segmented pneumonia infection regions to refine the attention for the network. Wang \emph{et al.} used an attention module to focus the learning on small targets to segment tumors in high-resolution T2-weighted MRI \cite{wang2019automatic}. Yan \emph{et al.} proposed an attention-based method for melanoma recognition, in which the attention maps are learned together with other network parameters \cite{yan2019melanoma}.

Note that these attention methods usually only produce one attention map for each image, as they mostly deal with images containing one single class. On the contrary, our task here is classifying multiple attributes and malignancy for each image. This task is different as an ideal model needs to put attention on various image regions to predict each task correctly. To solve this, we propose a novel attribute-specific multi-attention scheme that can adapt to different attributes freely. Experiment results show that this design can help boost the model performance and robustness.

\section{Methodology}
The proposed multi-attribute attention network (MAA-Net) includes three main parts:  a backbone, an attribute branch, and an attention branch. The backbone extracts the original input images to the high level features. The attribute branch learns to predict the nodular attributes and the nodule malignancy. The attention branch guides the attention of the model to further improve classification accuracy. The detailed design of the architecture is explained in the subsequent sections.

\subsection{An End-to-End Interpretable Diagnosis Framework}
Figure \ref{fig2} offers a schematic of the overall framework. An original image is first processed by the backbone $f(I)$, which maps the inputs $I$ to high-level features. In specific, ResNet50 is adopted. The extracted features are then fed into the attention branch to generate suitable attention heatmaps for each task. These heatmaps are then multiplied with the obtained feature maps and fed into the attribute branch.%In specific, ResNet-152 is adopted as the backbone. 

The attribute branch consists of several paralleled streams, corresponding to each desired nodular attribute, i.e. the shape, calcification, ratio, boundary, margin, and echo uniformity. Each stream has three residual blocks (3x3 convolutions, the first block has a stride of 2) to learn attribute specific features, a global average pooling (GAP) layer to summarize those features and a fully connected (FC) layer to yield the final prediction. To imitate the clinician diagnosis process, the model concatenates the six attribute-specific FC layers together to predict the nodule malignancy (shown as `cancer' branch in Fig. \ref{fig2}). By this way, the proposed model explicitly imitating the clinical work-flow in the `diagnosis-making' process. Note that no additional information is used except for the features of the nodular attributes, reinforcing the model interpretability.  

The ground truth of the model is a 1D vector, denoted as $L=\left[l_{1}, l_{2}, \ldots, l_{c}\right]$ in which $l_{c} \in\{0,1\}$, $\mathrm{C}=7$, $l_{c}$ presents the probability of being malignancy or belonging to a certain class of the corresponding attribute. A sigmoid layer is added to normalize the predicted output $p_{attr}(c \mid I)$ by:

\begin{equation}\tilde{p}_{attr}(c \mid I)=\frac{1}{1+\exp \left(-p_{attr}(c \mid I)\right)}\end{equation}
, where $I$ is the input image, $\tilde{p}_{attr}(c \mid I)$ presents the probability score of I belonging to the $c^{t h}$ class, $ c \in\{1,2, \ldots, C\} .$ We optimize the parameter of the attribute branch using the binary cross-entropy loss (BCE) loss:
\begin{equation}\begin{array}{r}
L_{attr}=-\frac{1}{C} \sum_{c=1}^{C} l_{c} \log \left(\tilde{p}_{attr}(c \mid I)\right)+\left(1-l_{c}\right) \log (1 \\
\left.-\tilde{p}_{attr}(c \mid I)\right)
\end{array}\end{equation}
, where $l_{c}$ is the ground truth label of the $c^{t h}$ class.

\subsection{Attribute-Adaptive Attention Scheme}
The human can direct their attention to the region of an image that is the most relevant to a specific task. For example, given the same US image, doctors usually look at the edge of the nodule to recognize whether its boundary is clear or blurry, while paying more attention to the internal region of the nodule to detect the existence of calcification. Inspired by this, we propose a flexible attribute-adaptive attention scheme to instruct the model to focus on different image regions in predicting various attributes.

\begin{figure}
    \centering
    \includegraphics[width = \columnwidth]{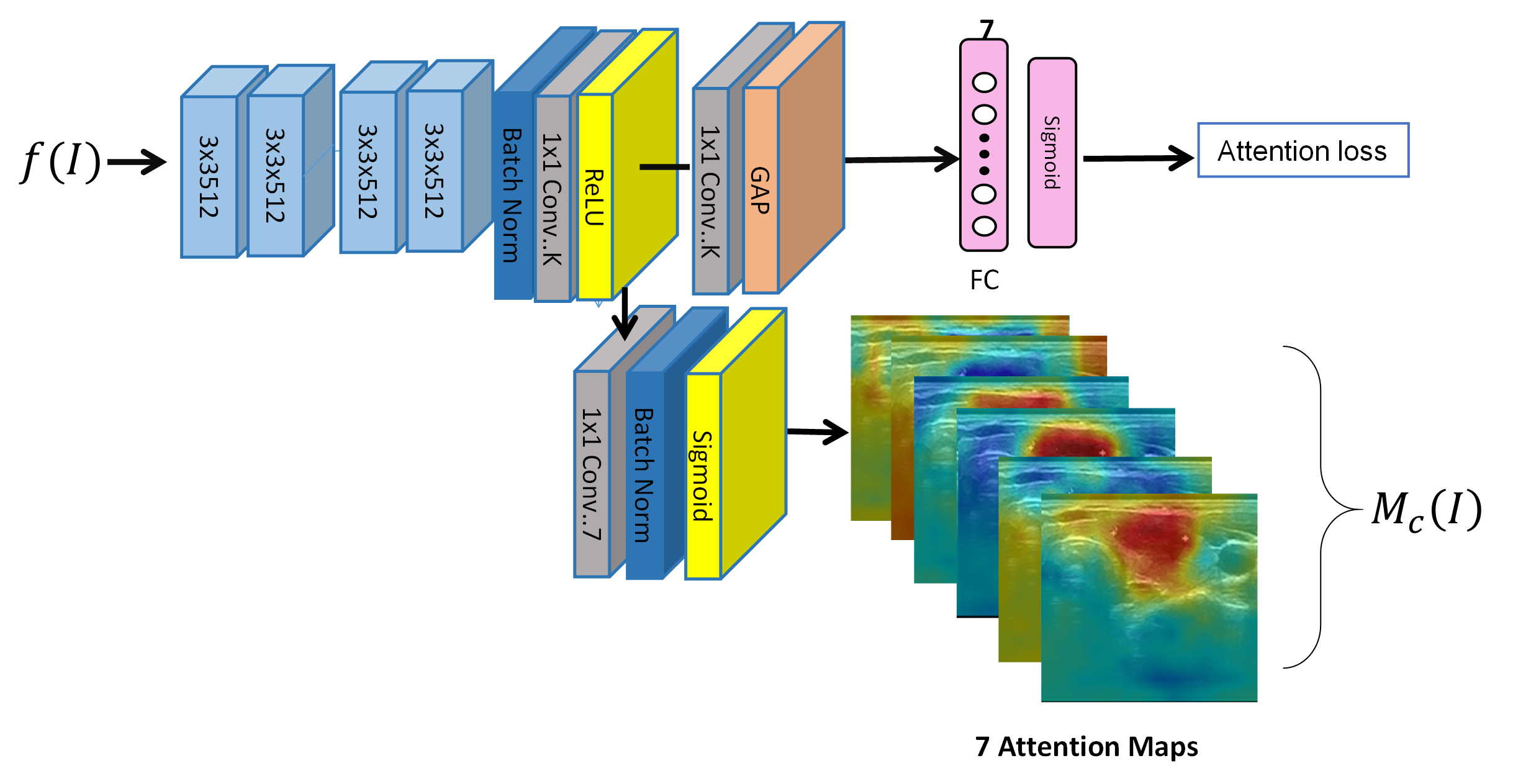}
    \caption{The structure of the attention branch. The global average pooling layer is represented by an orange cube. The light blue cubes represent the 3x3 convolutional block, the dark blue cube represents the batch-normalization layer, the yellow cube represents the ReLU or sigmoid function, and the gray cube represents the 1x1 convolutional layer. }
    \label{fig3}
\end{figure}

The structure of the attention branch is illustrated in Fig. \ref{fig3}. The extracted features are convoluted by 3x3 kernels with a stride of 1. A 1x1 convolution is used to aggregate the feature maps into a spatial size of 7x14x14. It is then sequentially normalized by the sigmoid function to get 7 different attention maps. Six of them are connected to the six attribute streams, respectively. The other attention map is used to impose an additional spatial constraint and will be explained in detail later. The 6 attribute-specific attention maps can highlight different regions of the image that is the most relevant to the corresponding attribute. By multiplying them with the original feature map, the model can put more emphasis on relevant information while avoiding distractions from the other. Formally, the filtered feature map $f_{c}\left(I\right)$ for the attribute $c^{t h}$ is calculated as:
\begin{equation}
f_{c}\left(I\right)=M_{c}\left(I\right) \cdot f\left(I\right),
\end{equation}
where $M_{c}\left(I\right)$ are the the attention maps. %$f\left(I\right)$ is the and multiplying the extracted feature $f\left(I\right)$ with the attention map $M_{c}\left(I\right)$
Note that, the attribute branch is also passed to a GAP layer and a sigmoid layer to add an auxiliary supervision for the malignancy and 6 attributes. This is then used to calculate the attention loss to supervise the attention branch directly, in specific:
\begin{equation}\begin{array}{r}
L_{attn}=-\frac{1}{C} \sum_{c=1}^{C} l_{c} \log \left(\tilde{p}_{attn}(c \mid I)\right)+\left(1-l_{c}\right) \log (1 \\
\left.-\tilde{p}_{attn}(c \mid I)\right),
\end{array}\end{equation}
where the $\tilde{p}_{attn}(c \mid I)$ presents the probability score of the attributes at the end of the attention branch. The $L_{attn}$ is used to put extra supervision for the attention maps.

\subsection{Spatial Attention}
Thyroid nodules have varying sizes and indefinite spatial locations within images. Existing methods usually crop or segment the regions of thyroid nodules first, which may lose context information and could lead to error-propagation. Instead, we propose to use the delineation of nodules to constrain the model's attention while retaining all available information. In specific, we force the model to attentively learn the prior spatial information of the 7th attention map (introduced in Section III.B) by matching with the mask of the original nodule (see Fig. \ref{fig2}).

To achieve this, the model penalizes the following loss:
\begin{equation}L_{\text {spatial}}=1-\frac{2 \sum_{i}^{N} p_{i} g_{i}}{\sum_{i}^{N} p_{i}+\sum_{i}^{N} g_{i}}\end{equation}
, where N denotes the number of pixels in the attention map, $p_i$ and $g_i$ represent pairs of corresponding pixel values of attention map and the delineation mask, respectively. The pixel value of $p_i$ and $g_i$ is normalized in range of [0;1]. Note that this additional loss is not directly imposed on any of the attribute streams as an ideal attention map may not be limited to the nodule region. For example, classifying the `echo uniformity' attribute requires knowledge of both the nodule region and all surrounding tissues. However, the spatial location and the shape of nodules are expected to be informative. Therefore, we generate an extra attention map for this loss which influences the generation of other attention maps through weight-sharing kernels. It can also serve as a localization head during the test. Later experiments validate that this spatially constrained attention map aligns well with the nodule, and it can increase the classification accuracy of nodular attributes as well.

The overall loss function of the whole framework is therefore defined as the combination of three. Formally: 
\begin{equation}L=w_{attr} L_{attr}+w_{attn} L_{attn}+w_{spatial} L_{spatial}\end{equation}
, where $w_{attr}$, $w_{attn}$, $w_{spatial}$ are the weights of the corresponding loss $L_{attr}$, $L_{attn}$, $L_{spatial}$. Empirically, the $w_{attr}$ is set to 1 for the main malignancy and attributes classification task. The $w_{attn}$, $w_{spatial}$ of attention loss and spatial loss are secondary for extra supervision and localization task and are set to 0.5 to not dominate the main task and better training convergence.

\section{Materials and Experiments}
\subsection{Dataset}
The thyroid dataset was collected from over 20 different sites and hospitals to capture various kinds of scans in a real clinical setting. In total, there are 4554 ultrasound images of thyroid nodules of 4554 patients with an age range from 9 to 82 years. The collected images contain only single nodule with the corresponding attributes and spatial mask information.  The ground truth classification labels were obtained based on the corresponding biopsy report, while other labels were annotated by doctors with over 20 years of experience.  The annotation process followed the TI-RADS guidelines [3] and the American Thyroid Association (ATA) guidelines [12]. The images were randomly shuffled and split to training (60\%), validation (20\%), and testing (20\%) sets.

\begin{figure*}
    \centering
    \includegraphics[width = \linewidth]{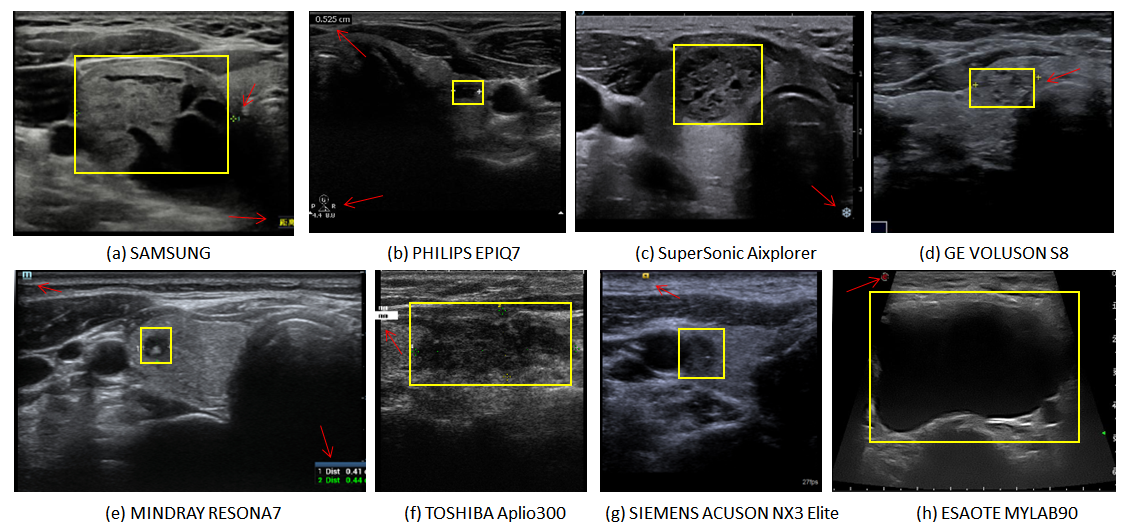}
    \caption{Examples of images collected using different machines. Note that these images have different contrast, intensity distribution, and the scanning window size. Meanwhile, the sizes and the locations of the nodules vary dramatically (shown in yellow boxes). Red arrows show some artifacts related to the different machines.}
    \label{fig4}
\end{figure*}

\begin{figure}
    \centering
    \includegraphics[width = \linewidth]{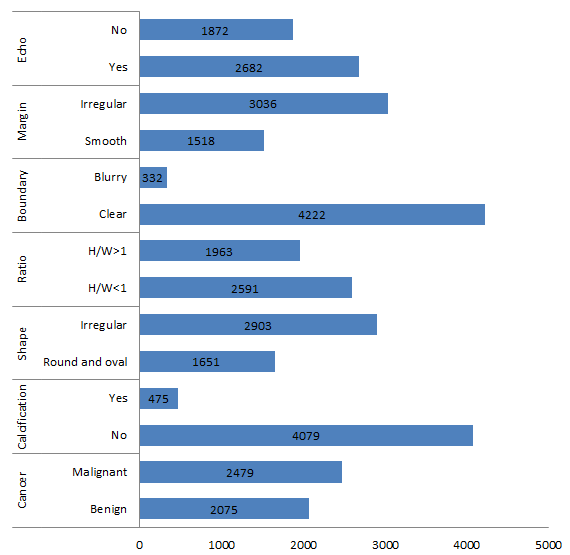}
    \caption{The distribution of nodule attributes and malignancy within the dataset.}
    \label{fig5}
\end{figure}

This rich dataset contains 2D ultrasound images acquired from a wide range of ultrasound machines e.g. Mindray DC-8, Mindray RESONA7, Philips-cx50, Philips EPIQ7, GE VOLUSON S8,  TOSHIBA Aplio 300, SIEMENS ACUSONS 2000, SIEMENS ACUSON NX3 Elite, and Esaote Mylab.  Note that the image contrast, overall intensity distribution, and the scanning window size were different.  Fig. \ref{fig4} shows some typical examples.  It also demonstrates that the sizes of the nodule have large variations (e.g. Fig. \ref{fig4}(b) and (h)). Moreover, their shape also varies (see Fig. \ref{fig4}(a) and (f)). The nodules can be cystic (Fig. \ref{fig4}(h)) or solid (Fig. \ref{fig4}(g)) and appear at different positions in the images. The total numbers of benign and malignant nodules and other attributes are shown in Fig. \ref{fig5}. It should be pointed out that this dataset was collected to approximate real-world clinical settings, while most existing medical datasets were collected using certain types of machines in one center/hospital. A robust algorithm that can perform well on such a challenging dataset is expected to perform similarly in a real-world application.

\subsection{Implementation Details}
The images are first converted to gray-scale in PNG format. To preserve the original aspect ratio of the nodules, the images are padded to a square shape before resizing to the same size of 224x224. No other preprocessing was performed. In training, data augmentation was carried out by resizing, random cropping, and random horizontal flipping. The ResNet50 with ImageNet pre-trained weights was used as the baseline. The network was trained for 150 epochs with a batch size of 20. The initial learning rate was 0.01 and divided by 10 after 20 epochs. The network was optimized using stochastic gradient descent (SGD) with a weight decay of 0.0001 and a momentum of 0.9. The MAA-Net was implemented with the Pytorch framework. We used a  Geforce GTX 1080 Ti GPU for the training and evaluation.

\subsection{Experiments}
\textbf{Comparison with state-of-the-art methods}. To demonstrate the efficacy of the proposed framework, we implemented some state-of-the-art methods using our dataset for a fair comparison. Intuitively, we implemented some classical classification methods such as ResNet50 \cite{he2016deep}, VGG-16 \cite{simonyan2014very},  and DenseNet201 \cite{huang2017densely}.  Furthermore, we also chose some of the most related works on DL-based thyroid nodule diagnosis \cite{buda2019management}\cite{guan2019deep}\cite{chi2017thyroid}. Note that these methods \cite{buda2019management},\cite{guan2019deep},\cite{chi2017thyroid} require an additional stage of localizing or pre-processing the ROIs using either human annotation or additional models. 

To investigate whether a single simple CNN model can combine the two tasks, we designed two additional comparison models which can solve the recognition and the classification task. In specific, the classical U-Net \cite{ronneberger2015u} model was chosen as the backbone for its simplicity. In `UNet-Simul' (Table \ref{tab1}), a classification head is added to the middle of the U-Net to predict the nodule malignancy. In the `UNet-2Stage' model, we investigated another way to combine the two tasks where the segmentation result is used to crop the nodule ROI and is then fed to another ResNet50 classification network. The experiment results are shown in Table \ref{tab1}.  

\textbf{Ablation study}. To evaluate the effectiveness of different components in our MAA-Net framework, we performed the following ablation experiments: 1) Feature extractor + single malignancy output (Baseline). 2) Feature extractor + Attribute branch (Baseline+Attr). 3) Feature extractor + Attribute branch+ Attention branch  (Baseline+Attr+Attn). 4) Feature extractor + Attribute branch+ Attention branch + Spatial loss  (the proposed MAA-Net).

\textbf{Cross-validation}. To further estimate if the proposed model can generalize to the whole dataset, a 5-fold cross-validation experiment was performed. The dataset was randomly partitioned equally into 5 sub-sets. Each cross-validation model was trained using the same hyper-parameters and data augmentation method described in Section IV-B.

\begin{table*}
\centering
\caption{Performance comparison of the state-of-the-art methods. `AUC' stands for area-under-curve, `Acc' stands for accuracy, `Hit rate' indicating whether the predicted nodule location correlates well with the ground truth. `Multi-stage' indicates whether the model is consists of two stages, `Pre-process'  indicates whether the method requires preprocessing,  and `Interp' stands for interpretability.}
\label{tab1}
\begin{tabular}{c|c|c|c|c|c|c}
\hline
Name                  & AUC Malignancy & Acc   & Multi-stage & Hit rate & Pre-process & Interp \\ \hline
ResNet50 \cite{he2016deep}     & 0.839       & 0.782 & no          & --       & --         & no     \\ \hline
VGG16 \cite{simonyan2014very}        & 0.794       & 0.754 & no          & --       & --         & no     \\ \hline
DenseNet201 \cite{huang2017densely}  & 0.840       & 0.785 & no          & --       & --         & no     \\ \hline
Buda et al. \cite{buda2019management}   & 0.842       & 0.786 & yes         & \textbf{0.910}     & no         & yes    \\ \hline
Guan et al. \cite{guan2019deep}  & 0.844       & 0.810  & yes          & --       & yes        & no     \\ \hline
Chi  et al. \cite{chi2017thyroid}  & 0.883       & 0.803 & no          & --       & yes        & no     \\ \hline
UNet-2Stage & 0.835       & 0.780  & yes         & 0.831     & no         & no     \\ \hline
UNet-Simul  & 0.771       & 0.750  & no        & 0.752     & no         & no     \\ \hline
MAA-Net (Ours)                  & \textbf{0.906}       & \textbf{0.836} & no          & 0.880     & no         & yes    \\ \hline
\end{tabular}
\end{table*}

\subsection{Evaluation Metrics}
To quantitatively evaluate the proposed MAA-Net model, we used the accuracy (Acc), F1score, specificity (Spec), precision (Prec), recall (Rec), and the area under the receiver operating characteristic curve (AUC) as performance metrics, which are defined as:
\begin{equation}
\begin{array}{c}
    Accuracy=\frac{T P+T N}{T P+F N+T N+F P},\\
    Specificity =\frac{T N}{T N+F P},\\
    Precision =\frac{T P}{T P+F P},\\
    Recall =\frac{T P}{T P+F N},\\
    F_{1 \text { Score }}=2 \frac{ Precision \times Recall }{ Precision + Recall },\\
    A U C=\int_{0}^{1} t_{\text {pr }}\left(f_{p r}\right) d f_{p r}=P(X 1>X 0),
\end{array}
\end{equation}

, where TP, FN, TN, and FP refer the number of true positive, false negative, true negative and false positive cases, respectively. $t_{pr}$ is the true positive rate, $f_{pr}$ is the false positive rate, and X0 and X1 are the confidence scores for a negative and positive instance, respectively.

\section{Result and Discussion}
\subsection{Comparison with the state-of-the-art methods}
The experiment results of the performance comparison between the proposed MAA-Net and other state-of-the-art methods are shown in Table \ref{tab1}. The column `Multi-stage' indicates whether the method has two stages, while the column `Pre-process' shows whether it requires pre-processing.  It shows that the proposed model achieved the highest accuracy (up to 0.836), AUC score (up to 0.906) on this challenging dataset while requiring no human intervention or multi-stage. This proves that the proposed MAA-Net can process original thyroid nodule images collected from various types of machines without pre-processing and is capable of dealing with variation of nodule location, size and appearance through learning the regions and context information.  

Table \ref{tab1} also shows that popular computer vision models such as ResNet50, VGG-16, Densenet201 (\cite{he2016deep},\cite{simonyan2014very},\cite{huang2017densely}) achieved relatively inferior performances. It might result from that these models were not specialized for this challenging original images with high variation characteristics of nodules. Among them, the DenseNet201 achieved higher accuracy, which may result from its rich connection among layers.

Note that some two-stage models achieved better accuracy and the AUC (e.g. \cite{buda2019management} Acc=0.786 and AUC=0.842). However, it is still lower than our methods. We conjecture that it might suffer from false detection in the first stage, which leads to incorrect cropping of the image and depleting the informative regions. The classification network cannot make accurate predictions on those images that might not even contain the targeted nodules. On the contrary, our method process the original images to predict in one shot therefore avoids similar errors. 

Other methods (e.g. \cite{guan2019deep}) avoid these kinds of errors by using manual pre-processing. We implemented these methods faithfully by using manually cropped images to train the corresponding models. According to Table \ref{tab1}, \cite{guan2019deep} achieved better accuracy than those two-staged methods (e.g. \cite{buda2019management}, UNet-2Stage), but was inferior to the MAA-Net. We conjecture that the surrounding tissues around the nodules may contain useful context information, such as whether the nodule tissue is solid or cystic. Therefore, the removal of the non-nodule image region might not cause information loss.

This finding also validates our model design which uses nodule delineation to constraint the model attention, rather than eliminating the background region completely. Note that the predicted attention maps can serve as a rough localization tool. We calculate the percentage of the predicted attention maps that have larger than 50\% overlap with the ground truth (shown as hit rate in Table \ref{tab1}). It can be seen that the predicted malignancy attention correlates well with the nodule location. It achieved a high hit rate (ours=0.88) similar to the separately trained localization models (e.g. that of \cite{buda2019management}=0.91).

It is also interesting to see that the UNet-Simul did not achieve comparable performance as expected. We reckon it might result from the competition between the two tasks (segmentation and classification) that lead to performance degradation in both. The UNet-2Stage avoids this competition by separating the two tasks. Table I shows that the latter scored higher classification performance compared to UNet-Simul, validating our previous conjectures. These experiments were also part of our inspiration to design a model that can exploit nodule location information without sacrificing classification accuracy.  The approach in \cite{chi2017thyroid} also achieved relatively good accuracy. However, it necessitates pre-processing to calibrate the scanning scale and remove possible artifacts. It then requires time and effort to generalize to images collected from new or un-seen types of US machines, which is frequently encountered in real-world settings. Meanwhile, only the proposed method and \cite{buda2019management} can predict the nodular attributes (marked ‘yes’ in the ‘Interp’ column), thus provide more interpretable results for the doctors. In the ablation study, we report the model accuracy in predicting these nodular attributes in detail.

\begin{table*}[]
\centering
\caption{Results of the ablation studies. `Malignancy' indicating the classification of nodule into malignant or benign class. Calcification (C),Shape (S), Ratio (R), Boundary (B), Margin (M), Echo (E).}
\label{tab2}
\begin{tabular}{c|cccccccc|ccccc}
\hline
\multirow{2}{*}{Model}                                           & \multicolumn{8}{c|}{AUC\_ROC}                                            & \multicolumn{5}{c}{Malignancy}      \\ \cline{2-14} 
                                                                 & Malignancy & Calci & Shape & Ratio & Boundary & Margin & Echo & Avg & Acc   & F1score & Spec  & Rec & Prec\\ \hline
\begin{tabular}[c]{@{}c@{}}Baseline\\   (ResNet50)\end{tabular} & 0.839  & --     & --     & --     & --        & --      & --         & --       & 0.782 & 0.813   & 0.738 & 0.848 & 0.781  \\ \hline
Baseline+C  & 0.865  & 0.642 & -- & -- & --  & --  & --    & --   & 0.789 & 0.829   & 0.635 & 0.911 & 0.761  \\ 
Baseline+S  & 0.875  & -- & 0.801 & -- & --  & --  & --    & --   & 0.792 & 0.831   & 0.635 & 0.916 & 0.761  \\ 
Baseline+R  & 0.864  & -- & -- & 0.764 & --  & --  & --    & --   & 0.791 & 0.822   & 0.694 & 0.866 & 0.783  \\ 
Baseline+B  & 0.859  & -- & -- & -- & 0.683  & --  & --    & --   & 0.777 & 0.823   & 0.588 & 0.927 & 0.741  \\
Baseline+M  & 0.869  & -- & -- & -- & --  & 0.798  & --    & --   & 0.789 & 0.813   & 0.748 & 0.822 & 0.805  \\
Baseline+E  & 0.883  & -- & -- & -- & --  & --  & 0.649    & --   & 0.785 & 0.797   & 0.821 & 0.757 & 0.843  \\ 
Baseline+S+R  & 0.858  & -- & 0.789 & 0.769 & --  & --  & --    & --   & 0.787 & 0.818   & 0.704 & 0.853 & 0.786  \\ 
Baseline+S+R+B  & 0.876  & -- & 0.801 & 0.799 & 0.686  & --  & --    & --   & 0.792 & 0.820   & 0.721 & 0.848 & 0.794  \\ \hline
Baseline+Attr                                                           & 0.866  & 0.877 & 0.769 & 0.754 & 0.707    & 0.830  & 0.724     & 0.790   & 0.795 & 0.821   & 0.738 & 0.841 & 0.802  \\ \hline
Baseline+Attr+Attn                                                       & 0.881  & 0.882 & 0.819 & 0.807 & \textbf{0.785}    & 0.871  & 0.740     & 0.826   & 0.804 & 0.824   & 0.784 & 0.819 & 0.828  \\ \hline
MAA-Net                                                          & \textbf{0.906}  & \textbf{0.892} & \textbf{0.827} & \textbf{0.837} & 0.755    & \textbf{0.877}  & \textbf{0.747}     & \textbf{0.834}   & \textbf{0.836} & \textbf{0.852}   & \textbf{0.824} & \textbf{0.846} & \textbf{0.859}  \\ \hline
\end{tabular}
\end{table*}

\begin{table*}[]
\centering
\caption{Average results of the 5-fold cross-validation experiments.}
\label{tab3}
\begin{tabular}{c|cccccccc|ccccc}
\hline
        & \multicolumn{8}{c|}{AUC\_ROC}                                            & \multicolumn{5}{c}{Malignancy}      \\ \hline
5-Fold  & Malignancy & Calci & Shape & Ratio & Boundary & Margin & Echo & Average & Acc   & F1score & Spec  & Rec & Prec \\ \hline
Average & 0.901  & 0.891 & 0.822 & 0.821 & 0.758    & 0.858  & 0.743     & 0.832   & 0.835 & 0.855   & 0.828 & 0.849 & 0.855  \\ \hline
STD     & 0.016  & 0.021 & 0.016 & 0.018 & 0.024    & 0.019  & 0.023     & 0.021   & 0.015 & 0.028   & 0.033 & 0.015  & 0.027 \\ \hline
\end{tabular}
\end{table*}

\begin{figure*}
    \centering
    \includegraphics[width = \linewidth]{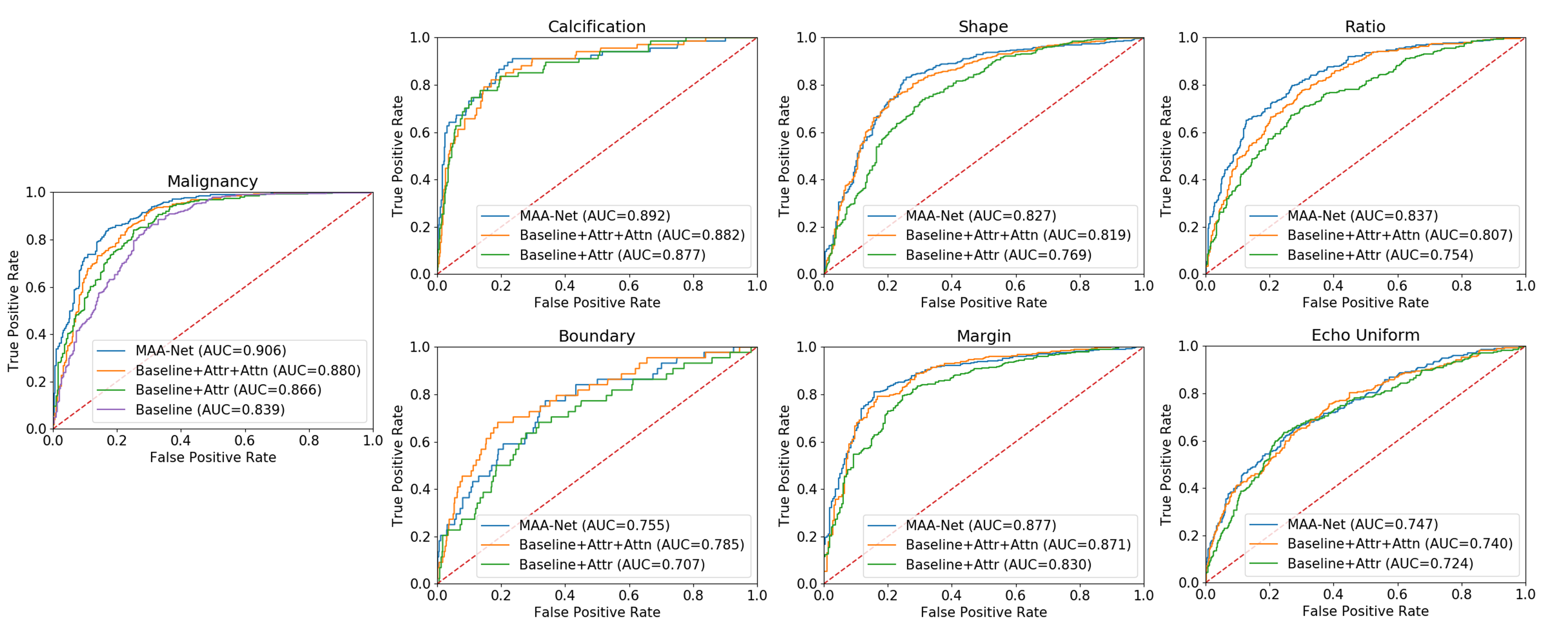}
    \caption{ROC curves of the ablation experiments over the attributes and malignancy predictions.}
    \label{fig6}
\end{figure*}

\subsection{Ablation Study}
\begin{figure*}
    \centering
    \includegraphics[width = \linewidth]{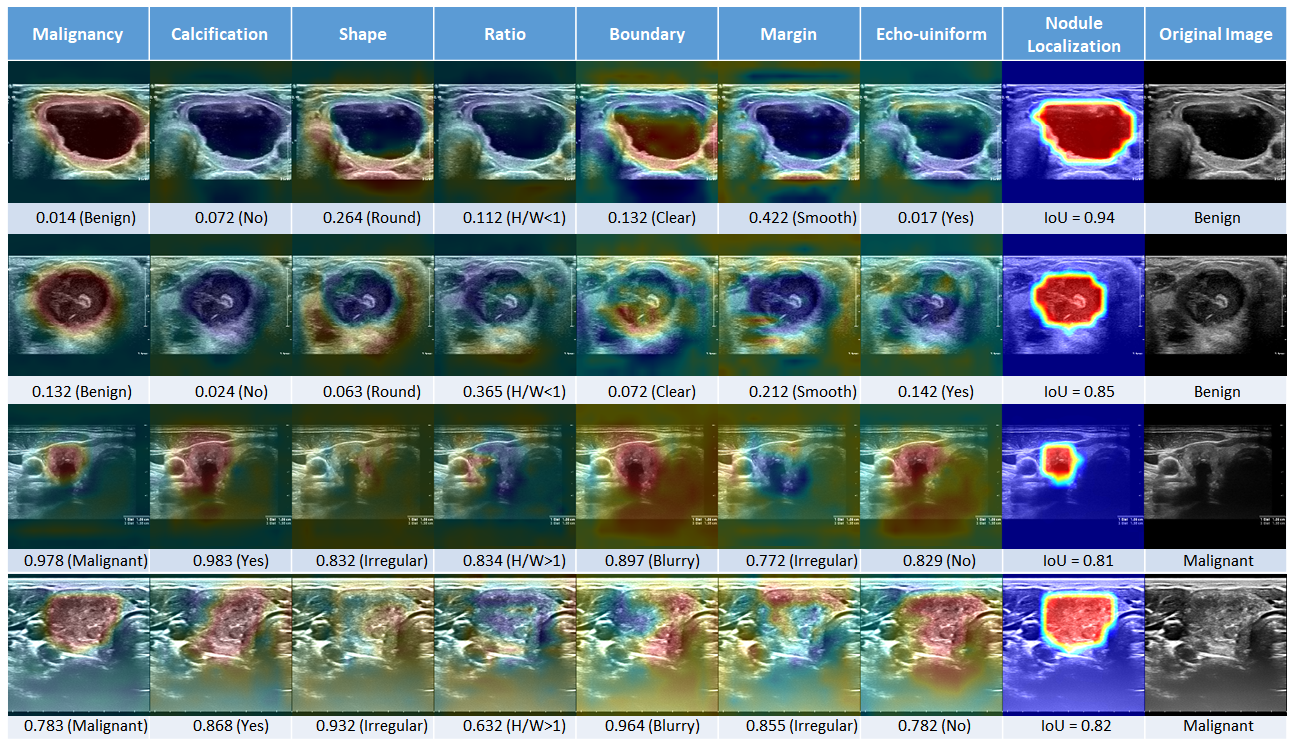}
    \caption{Visualization examples for attributes and malignancy of thyroid nodule. The heatmaps are rescaled to the same size and superimposed on the original images for better visualization. The prediction probabilities are shown to help doctors understand the diagnosis. }
    \label{fig7}
\end{figure*}

Table \ref{tab2} reports the results of the ablation study. It shows that adding the attribute branch with single attribute to the baseline (Baseline+C,S,R,B,M,E) can help to improve the malignancy classification accuracy. In addition, combining some attributes can not only improve the malignancy prediction but also simultaneously improve the prediction of attributes itself. As a result, combining all six attributes together (Baseline+Attr) help the baseline get the most improvement to accuracy of 0.795. It proves that the extra knowledge of the nodular attributes is indeed beneficial to improve the final malignancy prediction. Meanwhile, adding the attention branch (Baseline+Attr+Attn) also played a positive role in boosting model performance (see Table \ref{tab2}). The AUC results of six attributes are improved substantially compared to the Baseline+Attr. As a result, the classification accuracy for malignancy increases from 0.795 to 0.804, and the average AUC increases from 0.790 to 0.826. We conjecture that predicting multiple attributes is complex and may require the model to concentrate on different regions or utilize different features. Therefore, the proposed multi-attention scheme enables such dynamic learning. Visualization results also prove that the attention branch highlighted various local regions to better discriminate different attributes (see Fig. \ref{fig7}). 

Finally, the adding of the spatial attention loss resulted in the full MAA-Net, which further improved the average AUC of attribute prediction to 0.834 and Acc to 0.836. The spatial loss give a significant improvement in our method. Due to the challenge of using the original images instead of cropping the nodule, we conjecture that the spatial loss can give the rough localization of the nodule area which benefits the prediction of malignancy.  In clinical routine, the localization is also a importance step to analyze the thyroid nodule. Compared with results showed in Table \ref{tab1}, we argue that this design is more efficient and less error-prone than the two-staged methods. Furthermore, this attention maps can be generated automatically during the test and provide a rough localization of the nodule. Specifically, the precision is improved to 0.859 showing that our method can have less false positive and prevent the unnecessary biopsy as we expected. 

The Fig. \ref{fig6} visualizes the ROC curves of the ablation experiments over the nodule malignancy and attributes predictions. We can see that the attention mechanism and spatial loss can consistantly improve the true positive rates and AUC of all the attributes showing the effectiveness of the proposed components. As a result, the AUC of the malignancy diagnosis of MAA-Net can be largely improved (0.906) compared with the baseline (0.839).

\subsection{Cross Validation}
To further investigate whether the model performs consistently on this challenging dataset, we carried out a 5-fold cross-validation experiment. Table \ref{tab3} reports the average accuracy of the MAA-Net model trained and tested using different folds.  It shows that the proposed model is effective for all partitions of the dataset. It is also interesting to see that the model scored higher accuracy in recognizing the existence of calcification while achieved lower accuracy in predicting whether the boundary is clear or not. These results coincide with clinical studies on the inter-observer consistency in labeling these attributes \cite{choi2010interobserver},\cite{hoang2018interobserver}.

\subsection{Visualization of the Attributes and Malignancy}
To better understand the reasoning of the diagnosis, we provide visualization of the attention heatmaps generated for recognizing the attributes and the malignancy. Fig.\ref{fig7} provides four different examples for comparison in terms of malignancy and attributes. The prediction decisions and corresponding probability of the attributes provide the useful information for the doctor to understand. For example, the first row image is diagnosed as benign with high confidence because it has high probability of no calcification, round shape and clear boundary. In contrast, the third row image is malignant because it has high probability of calcification, H/W ratio $>$ 1, irregular shape and margin. These information consist with the TI-RADS guidelines \cite{grant2015thyroid}. Generally, the malignant attributes in row 3 and 4 have more high probability than benign cases in row 1,2 showing the better discrimination ability of model.

The two right-most columns show the original US images and the heatmaps generated to provide nodule location information. Note that the nodule regions are highlighted consistently thanks to the spatial attention loss, despite their contrasting sizes. Meanwhile, the column 1-7 are heatmaps correspond to the targeted nodular malignancy and attributes. It shows that the model concentrates on varying parts of the image to recognize different attributes. For instance, the `Calcification' and the `Echo uniform' heatmaps learn the region inside the nodule. The `Boundary', `Margin' heatmaps highlight the pixels around the nodule edge. This finding echoes with the previous conjecture that the model should focus on different image regions to better classify different attributes. Our attribute-based multi-attention scheme suits better to our multi-label task than the popular single attention scheme.

\section{Conclusion}
This paper proposed a novel MAA-Net framework for interpretable thyroid nodules diagnosis in ultrasound images. The design of the framework is inspired by clinical guidelines and human behaviors during diagnosis. This end-to-end approach uses an attribute branch to learn important clinical attributes and predicting the nodule malignancy based on those learned features. It is also equipped with a multi-attention scheme that can adapt to different attributes freely. The framework leverages the attention mechanism to guide the prediction, rather than cutting images based on human-intervention or pre-processing. Validation experiments showed that the proposed framework outperformed other state-of-the-art methods in malignancy classification and provided predictions of nodular attributes to explain its reasoning.

\bibliographystyle{IEEEtran}
\bibliography{maanet}

% Generated by IEEEtran.bst, version: 1.14 (2015/08/26)
\begin{thebibliography}{10}
\providecommand{\url}[1]{#1}
\csname url@samestyle\endcsname
\providecommand{\newblock}{\relax}
\providecommand{\bibinfo}[2]{#2}
\providecommand{\BIBentrySTDinterwordspacing}{\spaceskip=0pt\relax}
\providecommand{\BIBentryALTinterwordstretchfactor}{4}
\providecommand{\BIBentryALTinterwordspacing}{\spaceskip=\fontdimen2\font plus
\BIBentryALTinterwordstretchfactor\fontdimen3\font minus
  \fontdimen4\font\relax}
\providecommand{\BIBforeignlanguage}[2]{{%
\expandafter\ifx\csname l@#1\endcsname\relax
\typeout{** WARNING: IEEEtran.bst: No hyphenation pattern has been}%
\typeout{** loaded for the language `#1'. Using the pattern for}%
\typeout{** the default language instead.}%
\else
\language=\csname l@#1\endcsname
\fi
#2}}
\providecommand{\BIBdecl}{\relax}
\BIBdecl

\bibitem{haugen20162015}
B.~R. Haugen, E.~K. Alexander, K.~C. Bible, G.~M. Doherty, S.~J. Mandel, Y.~E.
  Nikiforov, F.~Pacini, G.~W. Randolph, A.~M. Sawka, M.~Schlumberger
  \emph{et~al.}, ``2015 american thyroid association management guidelines for
  adult patients with thyroid nodules and differentiated thyroid cancer: the
  american thyroid association guidelines task force on thyroid nodules and
  differentiated thyroid cancer,'' \emph{Thyroid}, vol.~26, no.~1, pp. 1--133,
  2016.

\bibitem{schnadig2018overdiagnosis}
V.~J. Schnadig, ``Overdiagnosis of thyroid cancer: is this not an ethical issue
  for pathologists as well as radiologists and clinicians?'' \emph{Archives of
  pathology \& laboratory medicine}, vol. 142, no.~9, pp. 1018--1020, 2018.

\bibitem{papini2002risk}
E.~Papini, R.~Guglielmi, A.~Bianchini, A.~Crescenzi, S.~Taccogna, F.~Nardi,
  C.~Panunzi, R.~Rinaldi, V.~Toscano, and C.~M. Pacella, ``Risk of malignancy
  in nonpalpable thyroid nodules: predictive value of ultrasound and
  color-doppler features,'' \emph{The Journal of Clinical Endocrinology \&
  Metabolism}, vol.~87, no.~5, pp. 1941--1946, 2002.

\bibitem{rahbari2010thyroid}
R.~Rahbari, L.~Zhang, and E.~Kebebew, ``Thyroid cancer gender disparity,''
  \emph{Future Oncology}, vol.~6, no.~11, pp. 1771--1779, 2010.

\bibitem{grant2015thyroid}
E.~G. Grant, F.~N. Tessler, J.~K. Hoang, J.~E. Langer, M.~D. Beland, L.~L.
  Berland, J.~J. Cronan, T.~S. Desser, M.~C. Frates, U.~M. Hamper
  \emph{et~al.}, ``Thyroid ultrasound reporting lexicon: white paper of the acr
  thyroid imaging, reporting and data system (tirads) committee,''
  \emph{Journal of the American college of radiology}, vol.~12, no.~12, pp.
  1272--1279, 2015.

\bibitem{choi2010interobserver}
S.~H. Choi, E.-K. Kim, J.~Y. Kwak, M.~J. Kim, and E.~J. Son, ``Interobserver
  and intraobserver variations in ultrasound assessment of thyroid nodules,''
  \emph{Thyroid}, vol.~20, no.~2, pp. 167--172, 2010.

\bibitem{hoang2018interobserver}
J.~K. Hoang, W.~D. Middleton, A.~E. Farjat, S.~A. Teefey, N.~Abinanti, F.~J.
  Boschini, A.~J. Bronner, N.~Dahiya, B.~S. Hertzberg, J.~R. Newman
  \emph{et~al.}, ``Interobserver variability of sonographic features used in
  the american college of radiology thyroid imaging reporting and data
  system,'' \emph{American Journal of Roentgenology}, vol. 211, no.~1, pp.
  162--167, 2018.

\bibitem{liu2019automated}
T.~Liu, Q.~Guo, C.~Lian, X.~Ren, S.~Liang, J.~Yu, L.~Niu, W.~Sun, and D.~Shen,
  ``Automated detection and classification of thyroid nodules in ultrasound
  images using clinical-knowledge-guided convolutional neural networks,''
  \emph{Medical image analysis}, vol.~58, p. 101555, 2019.

\bibitem{buda2019management}
M.~Buda, B.~Wildman-Tobriner, J.~K. Hoang, D.~Thayer, F.~N. Tessler, W.~D.
  Middleton, and M.~A. Mazurowski, ``Management of thyroid nodules seen on us
  images: deep learning may match performance of radiologists,''
  \emph{Radiology}, vol. 292, no.~3, pp. 695--701, 2019.

\bibitem{guan2019deep}
Q.~Guan, Y.~Wang, J.~Du, Y.~Qin, H.~Lu, J.~Xiang, and F.~Wang, ``Deep learning
  based classification of ultrasound images for thyroid nodules: a large scale
  of pilot study,'' \emph{Annals of Translational Medicine}, vol.~7, no.~7,
  2019.

\bibitem{chi2017thyroid}
J.~Chi, E.~Walia, P.~Babyn, J.~Wang, G.~Groot, and M.~Eramian, ``Thyroid nodule
  classification in ultrasound images by fine-tuning deep convolutional neural
  network,'' \emph{Journal of digital imaging}, vol.~30, no.~4, pp. 477--486,
  2017.

\bibitem{keramidas2008thyroid}
E.~G. Keramidas, D.~K. Iakovidis, D.~Maroulis, and N.~Dimitropoulos, ``Thyroid
  texture representation via noise resistant image features,'' in \emph{2008
  21st IEEE International Symposium on Computer-Based Medical Systems}.\hskip
  1em plus 0.5em minus 0.4em\relax IEEE, 2008, pp. 560--565.

\bibitem{ardakani2018predictive}
A.~A. Ardakani, A.~Mohammadzadeh, N.~Yaghoubi, Z.~Ghaemmaghami, R.~Reiazi,
  A.~H. Jafari, S.~Hekmat, M.~B. Shiran, and A.~Bitarafan-Rajabi, ``Predictive
  quantitative sonographic features on classification of hot and cold thyroid
  nodules,'' \emph{European Journal of Radiology}, vol. 101, pp. 170--177,
  2018.

\bibitem{tsantis2009morphological}
S.~Tsantis, N.~Dimitropoulos, D.~Cavouras, and G.~Nikiforidis, ``Morphological
  and wavelet features towards sonographic thyroid nodules evaluation,''
  \emph{Computerized Medical Imaging and Graphics}, vol.~33, no.~2, pp. 91--99,
  2009.

\bibitem{singh2012ultra}
N.~Singh and A.~Jindal, ``Ultra sonogram images for thyroid segmentation and
  texture classification in diagnosis of malignant (cancerous) or benign
  (non-cancerous) nodules,'' \emph{Int. J. Eng. Innov. Technol}, vol.~1, pp.
  202--206, 2012.

\bibitem{acharya2014computer}
U.~R. Acharya, S.~V. Sree, M.~M.~R. Krishnan, F.~Molinari, W.~Ziele{\"Y}nik,
  R.~H. Bardales, A.~Witkowska, and J.~S. Suri, ``Computer-aided diagnostic
  system for detection of hashimoto thyroiditis on ultrasound images from a
  polish population,'' \emph{Journal of Ultrasound in medicine}, vol.~33,
  no.~2, pp. 245--253, 2014.

\bibitem{raghavendra2017fusion}
U.~Raghavendra, U.~R. Acharya, A.~Gudigar, J.~H. Tan, H.~Fujita, Y.~Hagiwara,
  F.~Molinari, P.~Kongmebhol, and K.~H. Ng, ``Fusion of spatial gray level
  dependency and fractal texture features for the characterization of thyroid
  lesions,'' \emph{Ultrasonics}, vol.~77, pp. 110--120, 2017.

\bibitem{ma2017cascade}
J.~Ma, F.~Wu, T.~Jiang, J.~Zhu, and D.~Kong, ``Cascade convolutional neural
  networks for automatic detection of thyroid nodules in ultrasound images,''
  \emph{Medical physics}, vol.~44, no.~5, pp. 1678--1691, 2017.

\bibitem{elshawi2019interpretability}
R.~Elshawi, M.~H. Al-Mallah, and S.~Sakr, ``On the interpretability of machine
  learning-based model for predicting hypertension,'' \emph{BMC medical
  informatics and decision making}, vol.~19, no.~1, p. 146, 2019.

\bibitem{carvalho2019machine}
D.~V. Carvalho, E.~M. Pereira, and J.~S. Cardoso, ``Machine learning
  interpretability: A survey on methods and metrics,'' \emph{Electronics},
  vol.~8, no.~8, p. 832, 2019.

\bibitem{zhang2021survey}
Y.~Zhang and Q.~Yang, ``A survey on multi-task learning,'' \emph{IEEE
  Transactions on Knowledge and Data Engineering}, 2021.

\bibitem{wang2018simultaneous}
P.~Wang, V.~M. Patel, and I.~Hacihaliloglu, ``Simultaneous segmentation and
  classification of bone surfaces from ultrasound using a multi-feature guided
  cnn,'' in \emph{International conference on medical image computing and
  computer-assisted intervention}.\hskip 1em plus 0.5em minus 0.4em\relax
  Springer, 2018, pp. 134--142.

\bibitem{murugesan2019psi}
B.~Murugesan, K.~Sarveswaran, S.~M. Shankaranarayana, K.~Ram, J.~Joseph, and
  M.~Sivaprakasam, ``Psi-net: Shape and boundary aware joint multi-task deep
  network for medical image segmentation,'' in \emph{2019 41st Annual
  International Conference of the IEEE Engineering in Medicine and Biology
  Society (EMBC)}.\hskip 1em plus 0.5em minus 0.4em\relax IEEE, 2019, pp.
  7223--7226.

\bibitem{roy2020deep}
S.~Roy, W.~Menapace, S.~Oei, B.~Luijten, E.~Fini, C.~Saltori, I.~Huijben,
  N.~Chennakeshava, F.~Mento, A.~Sentelli \emph{et~al.}, ``Deep learning for
  classification and localization of covid-19 markers in point-of-care lung
  ultrasound,'' \emph{IEEE transactions on medical imaging}, vol.~39, no.~8,
  pp. 2676--2687, 2020.

\bibitem{zhou2016learning}
B.~Zhou, A.~Khosla, A.~Lapedriza, A.~Oliva, and A.~Torralba, ``Learning deep
  features for discriminative localization,'' in \emph{Proceedings of the IEEE
  conference on computer vision and pattern recognition}, 2016, pp. 2921--2929.

\bibitem{selvaraju2017grad}
R.~R. Selvaraju, M.~Cogswell, A.~Das, R.~Vedantam, D.~Parikh, and D.~Batra,
  ``Grad-cam: Visual explanations from deep networks via gradient-based
  localization,'' in \emph{Proceedings of the IEEE international conference on
  computer vision}, 2017, pp. 618--626.

\bibitem{ribeiro2016should}
M.~T. Ribeiro, S.~Singh, and C.~Guestrin, ``" why should i trust you?"
  explaining the predictions of any classifier,'' in \emph{Proceedings of the
  22nd ACM SIGKDD international conference on knowledge discovery and data
  mining}, 2016, pp. 1135--1144.

\bibitem{ouyang2020dual}
X.~Ouyang, J.~Huo, L.~Xia, F.~Shan, J.~Liu, Z.~Mo, F.~Yan, Z.~Ding, Q.~Yang,
  B.~Song \emph{et~al.}, ``Dual-sampling attention network for diagnosis of
  covid-19 from community acquired pneumonia,'' \emph{IEEE Transactions on
  Medical Imaging}, 2020.

\bibitem{wang2019automatic}
G.~Wang, J.~Shapey, W.~Li, R.~Dorent, A.~Demitriadis, S.~Bisdas, I.~Paddick,
  R.~Bradford, S.~Zhang, S.~Ourselin \emph{et~al.}, ``Automatic segmentation of
  vestibular schwannoma from t2-weighted mri by deep spatial attention with
  hardness-weighted loss,'' in \emph{International Conference on Medical Image
  Computing and Computer-Assisted Intervention}.\hskip 1em plus 0.5em minus
  0.4em\relax Springer, 2019, pp. 264--272.

\bibitem{yan2019melanoma}
Y.~Yan, J.~Kawahara, and G.~Hamarneh, ``Melanoma recognition via visual
  attention,'' in \emph{International Conference on Information Processing in
  Medical Imaging}.\hskip 1em plus 0.5em minus 0.4em\relax Springer, 2019, pp.
  793--804.

\bibitem{he2016deep}
K.~He, X.~Zhang, S.~Ren, and J.~Sun, ``Deep residual learning for image
  recognition,'' in \emph{Proceedings of the IEEE conference on computer vision
  and pattern recognition}, 2016, pp. 770--778.

\bibitem{simonyan2014very}
K.~Simonyan and A.~Zisserman, ``Very deep convolutional networks for
  large-scale image recognition,'' \emph{arXiv preprint arXiv:1409.1556}, 2014.

\bibitem{huang2017densely}
G.~Huang, Z.~Liu, L.~Van Der~Maaten, and K.~Q. Weinberger, ``Densely connected
  convolutional networks,'' in \emph{Proceedings of the IEEE conference on
  computer vision and pattern recognition}, 2017, pp. 4700--4708.

\bibitem{ronneberger2015u}
O.~Ronneberger, P.~Fischer, and T.~Brox, ``U-net: Convolutional networks for
  biomedical image segmentation,'' in \emph{International Conference on Medical
  image computing and computer-assisted intervention}.\hskip 1em plus 0.5em
  minus 0.4em\relax Springer, 2015, pp. 234--241.

\end{thebibliography}

\vfill

\end{document}